\documentclass[aps,pra,amssymb,superscriptaddress,nofootinbib,twocolumn]{revtex4-1}
\usepackage{graphicx}
\usepackage{graphicx}
\usepackage{bm}
\usepackage{amsmath}
\usepackage{amssymb}

\begin{document}
\title{Encoding relativistic  potential dynamics into free evolution}
\author{C. Sab\'in}
\email{csl@iff.csic.es}
\affiliation{Instituto de F\'{\i}sica Fundamental, CSIC,
  Serrano 113-B, 28006 Madrid, Spain}
\author{J. Casanova}
\affiliation{Departamento de Qu\'{\i}mica F\'{\i}sica, Universidad del Pa\'{\i}s Vasco UPV/EHU, Apdo.\ 644, 48080 Bilbao, Spain}
\author{J. J. Garc\'ia-Ripoll}
\affiliation{Instituto de F\'{\i}sica Fundamental, CSIC,
  Serrano 113-B, 28006 Madrid, Spain}
\author{L. Lamata}
\affiliation{Departamento de Qu\'{\i}mica F\'{\i}sica, Universidad del Pa\'{\i}s Vasco UPV/EHU, Apdo.\ 644, 48080 Bilbao, Spain}
\author{E. Solano}
\affiliation{Departamento de Qu\'{\i}mica F\'{\i}sica, Universidad del Pa\'{\i}s Vasco UPV/EHU, Apdo.\ 644, 48080 Bilbao, Spain}
\affiliation{IKERBASQUE, Basque Foundation for Science, Alameda Urquijo 36, 48011 Bilbao, Spain}
\author{J. Le\'on}
\affiliation{Instituto de F\'{\i}sica Fundamental, CSIC,
  Serrano 113-B, 28006 Madrid, Spain}
  
\begin{abstract}
We propose a method to simulate a Dirac or Majorana equation evolving under particular potentials with the use of the corresponding free evolution, while the potential dynamics is encoded in a static transformation upon the initial state. We extend our results to interacting two-body systems.
\end{abstract}

\maketitle

\section{Introduction}
The last years have witnessed an increasing interest in simulating dynamics coming from the relativistic quantum mechanics realm in different physical systems, e.g. in trapped ions~\cite{reviewrqm}, optical lattices~\cite{weitz, esslinger}, and quantum photonics~\cite{szameit}. Striking theoretical predictions related with the Dirac equation \cite{thaller} like \textit{Zitterbewegung}  and Klein paradox \cite{klein} have been observed in these simulations. In particular, the proposal of simulation in one trapped ion \cite{lucas}  of  the free single-particle one-dimensional (1D) Dirac equation has been successfully implemented in the lab \cite{naturekike}. This is also the case for the single-particle 1D  Dirac equation with some external potentials under which the particle exhibits Klein tunneling \cite{jorge1}, where the experimental implementation involved two ions \cite{jorge2}.

In reference~\cite{MajoranaSimulation:2010}, it was proposed that the free Majorana equation \cite{majoranaeq1, majoranaeq2} and unphysical operations like complex conjugation, charge conjugation, and time reversal  can also be simulated with two trapped ions. Besides, two-body Dirac equations have been the subject of recent theoretical research \cite{bermudez, reviewrqm}. In general, simulations of single-particle equations with external potentials or many-body equations  for interacting systems seem to be much more demanding than free equations.

In this work, we show that the same setups employed for the simulations of the free single-particle Dirac and Majorana equations can also be used for simulations of these equations with the addition of  a broad class of potentials. This is based in the following concept, which is the main result of this paper: any state which is a solution of the Dirac or Majorana equation with one of these potentials can be related through a static transformation with a solution of the free corresponding equation. Accordingly, in order to simulate the dynamics of a given state under certain potential, which could be cumbersome in some situations, all that is needed is to initialize the appropriate state and let it evolve under the free equation, which is often easier to implement. The method includes scalar and spinorial position-dependent potentials, different in the Dirac and Majorana equations. In general, the transformation does not leave the probability density unchanged but this happens in some particular cases, showing us an additional interesting feature, namely, under certain potentials the particle behaves as a free particle. In other cases, we show that the method works approximately in some regions of space or in some parameter ranges. In particular, a massive-particle dynamics can be simulated in this way with a massless-particle equation. Finally, we  show how to extend our results to two-body situations.

Although our method works for any dimension and representation, we focus here in the 1D case which has a direct experimental connection with  trapped-ion experiments  simulating relativistic quantum dynamics. We use also the particular representations employed in these experiments. The techniques developed in this work are also valid for 3D but the particular results obtained in that case must be the focus of further research.

\section{One particle systems}

We  will consider  the following 1D Dirac equation in natural units ($\hbar=c=1$),
\begin{equation}
i\dot{\psi}=-i\,\sigma_x \psi '+[\sigma_z\,m+V(x)]\psi, \label{eq:dirac}
\end{equation}
where $\dot{}$ and $'$ denote time and space partial derivatives respectively. In the same representation, the Majorana equation reads
\begin{equation}
i\dot{\psi}=-i\sigma_x\, \psi'-i\sigma_y\,m\psi^*+V(x)\psi. \label{eq:majorana}
\end{equation}
Moreover, in 1D a general potential can be written as \cite{jorge1}
\begin{equation}
V(x)= f_1(x)+ f_2(x)\,\sigma_z+f_3(x)\,\sigma_y+f_4(x)\,\sigma_x. \label{eq:generalpotential}
\end{equation}
Note that in the case of Eq. (\ref{eq:dirac}), we could consider mass-like potentials taking $f_2(x)=m$.

\subsection{Majorana equation} We start from Eq. (\ref{eq:majorana}) where the potential is given by Eq.
(\ref{eq:generalpotential}). We will analyze a set of sufficient conditions under which a  transformation of the form
\begin{equation}
\psi= \mathcal{U}(x) \phi, \label{eq:generaltransformation} 
\end{equation}
 where
\begin{equation}
\mathcal{U}(x)=
e^{-i\,F_1(x)\sigma_x-i\,F_2(x)\sigma_y-i\,F_3(x)\sigma_z-i\,F_4(x)},\label{eq:phase}
\end{equation}
 can convert  Eq. (\ref{eq:majorana}) into the corresponding free Majorana equation for $\phi$. (Notice that
 $e^A\,e^B\ne e^{A+B}$ unless $[A,B]=0$, which in general is not the case here.) To this end, we first notice that after applying Eq. (\ref{eq:generaltransformation}) the LHS of Eq. (\ref{eq:majorana}) becomes
\begin{equation}
i \mathcal{U}(x)\dot{\phi} ,\label{eq:lhs}
\end{equation}
 while the RHS transforms into
 \begin{equation}
-i \sigma_x \big[\mathcal{U}(x)
\phi'+\mathcal{U}(x)' \phi\big]-i m \sigma_y  \mathcal{U}(x)^* \phi^*+ V(x)\mathcal{U}(x) \phi, \label{eq:rhs1}
\end{equation}
We now impose
\begin{equation}
 i \sigma_x\mathcal{U}(x)' = V(x)\mathcal{U}(x).\label{eq:rempot1}
\end{equation}
In the cases where the derivative of the exponent of $\mathcal{U}$ commutes with $\mathcal{U}$, i.e., restricting $\mathcal{U}$ in Eq.~(\ref{eq:phase}) to 
\begin{equation}\label{particular}
\mathcal{U}(x)=e^{-i F_j(x)\sigma_j - i F_4(x)},
\end{equation}
with $\sigma_j = \sigma_x, \sigma_y, \sigma_z$ for $j=1,2,3$ respectively, we can write Eq.~(\ref{eq:rempot1}) as
\begin{equation}
 \sigma_x \big[F'_j(x)\sigma_j + F'_4(x)\big]\mathcal{U}(x) = V(x)\mathcal{U}(x).
\end{equation}

This amounts to removing the potential from the dynamical equation for $\phi$ choosing properly the relationships between the functions $F_j(x)$ and   $f_j(x)$ in Eqs.~(\ref{eq:generalpotential}), (\ref{eq:phase}). Specially interesting is the case
 \begin{eqnarray}
F_1 '(x) = f_1(x), &&F_4'(x) = f_4(x),  \nonumber \\  
F_2(x) = F_3(x) &=& f_2(x) = f_3(x) = 0, \label{eq:generalproperty2}
\end{eqnarray}
that represents an electromagnetic potential acting on the Majorana particle.
Under this condition,  we will now set the commutation rules of $\mathcal{U}$ with the Pauli matrices,
\begin{equation} 
\sigma_x  \mathcal{U} = \widetilde{\mathcal{U}}\sigma_x
,\, \mbox{and}\; \sigma_y \mathcal{U}^* = \widetilde{\mathcal{U}}\sigma_y, \label{conds1}
\end{equation} 
while the first
is simply the definition of $\widetilde{\mathcal{U}}$, the second gives $\mathcal{U}^*$ in terms of $\widetilde{\mathcal{U}}$, i.e.
\begin{equation} 
\widetilde{\mathcal{U}} = \sigma_x  \mathcal{U} \sigma_x,\, \mbox{and} \;  \mathcal{U}^* = \sigma_y
\widetilde{\mathcal{U}} \sigma_y, \label{conds2}
\end{equation}
which implies that $F_1$ has to be real and $F_4$
imaginary. In this case, the Majorana equation (\ref{eq:majorana}) becomes
\begin{equation} 
i\mathcal{U}\dot{\phi}
=\widetilde{\mathcal{U}} (-i \sigma_x \phi'-i m \sigma_y \phi^*),\label{majoranamod}
\end{equation}
where $\mathcal{U}=\widetilde{\mathcal{U}}$ and $\phi$ satisfies a free Majorana equation. 

That is, if we want to simulate  the dynamics of a particle in the state $\psi(x,t)$ under the 1D  Majorana equation with a potential of the form 
\begin{equation}
V(x)= f_1(x)+f_4 (x) \sigma_x, 
\end{equation}
we only have to prepare the initial state $\phi(x,t = 0)$ which is related with $\psi(x, t=0)$ through 
\begin{equation}
\psi(x,t=0)=e^{-i\,F_1(x)\sigma_x-i\,F_4(x)}\,\phi(x,t=0),
\end{equation}
and then let the system evolve under the free Hamiltonian of Eq.~(\ref{majoranamod}).

Since $F_1$ is real and $F_4$ imaginary, we have the following relationship between the probability densities:
\begin{equation}
|\psi(x,t)|^2=e^{-2\,i\,F_4(x)}\,|\phi(x,t)|^2.\label{eq:probabilities}
\end{equation}
Accordingly, the probability density observed for $\phi$ can be easily related with the simulated one for $\psi$. Analogous relations can also be derived, for instance, for expectation values of observables. 

Of particular interest is the case in which $F_4(x)=0$, where the potential does not include the term with  $f_4(x)$, because the probability density is the same for $\psi$ and $\phi$. Therefore, the probability density of a state $\psi(x, t)$ under the potential $V(x)=f_1(x)$, is always the same as the one of the transformed  free state $\phi(x, t)$. We shall illustrate this with an example below.

\subsection{Dirac equation} We now analyze the case of the 1D Dirac equation of Eq. (\ref{eq:dirac}), assuming that we have a massive system $m\neq0$. Using the same techniques as in the Majorana case, we can eliminate a potential of the form 
\begin{equation}
V (x)= f_4(x) \sigma_x \label{eq:potdir} 
\end{equation}
from Eq.~(\ref{eq:dirac}), with the transformation 
\begin{equation}
\psi = e^{-iF_4(x)}\phi, \label{eq:transpotdir}
\end{equation}
 where 
 \begin{equation}
 F'_4(x) = f_4(x). \label{eq:condtranspotdir}
 \end{equation}
 The resulting dynamical equation for $\phi$ is 
\begin{equation}\label{dinphi}
i\dot{\phi} = -i \sigma_x \phi' +  \sigma_z m \phi,
\end{equation}
and the relation between $|\psi|$ and $|\phi|$ is given by 
\begin{equation}
|\psi| = e^{2{\rm Im}[F_4(x)] }|\phi|,
\end{equation}
with ${\rm Im}[F_4(x)] $ the imaginary part of $F_4(x)$. Note that with this method we are allowed to simulate the presence of complex potentials in the Dirac dynamics through the Hermitian free equation (\ref{dinphi}). This may be useful to study weak dissipative processes. 

\subsection{Non-relativistic limit}  The case of a Dirac potential with only $f_4\neq0$, as is the case for the minimal coupling with an electromagnetic field, can be solved exactly and is related to the U(1) gauge invariance of Quantum Electrodynamics (QED). This also shows the partial relationship between our transformations and the standard gauge transformations of quantum field theories. This kind of potential gives rise, in the non-relativistic limit $p\ll m$, to a 1D Pauli-Schr\"odinger equation (without scalar potential), with Hamiltonian $H=[p-f_4(x)]^2/2m$. Since the non-relativistic limit amounts to the dispersive limit of the corresponding simulating trapped-ion equation \cite{lucas}, this shows that our method can also be applied to the simulation of non-relativistic dynamics. In the QED case, the vector potential considered in this example would be irrotational, such that it does not contribute to the interaction. However, we consider here a more general formalism, in which an arbitrary potential $V(x)= f_4(x)\,\sigma_x$ can be eliminated by the phase transformation.

As we have seen in Eqs. (\ref{eq:potdir})-(\ref{dinphi}), we will have a potential $V(x)=f_4(x)\,\sigma_x,$ that can be eliminated from the dynamical equation for $\psi$ through the transformation $\psi=e^{-i\,F_4(x)}\phi$
that turns Eq. (\ref{eq:dirac}) into $ i\dot{\phi} =-i\sigma_x \phi '\,+\sigma_z\,m\phi.\label{eq:minimallycoupledresultdirac}$
Thus the potential can be exactly eliminated.

Some deviations to the exact equivalences between free and potential-dependent dynamics  presented here can be considered.  For example, in the Dirac case a potential with $f_1(x)=f_3(x)=0$,
\begin{equation}
V(x)=f_2(x)\,\sigma_z+f_4(x)\,\sigma_x, \label{eq:generalpotentialdirac}
\end{equation}
can be erased from the dynamical equation of $\psi$ through the transformation
\begin{equation}
\psi=e^{-i\,F_2(x)\sigma_y-i\,F_4(x)}\phi,\label{eq:generaltransformationdirac}
\end{equation}
where 
\begin{eqnarray}
F_2 '(x) = -if_2(x), &&F_4'(x) = f_4(x),  \nonumber \\  
F_1(x) = F_3(x) &=& f_1(x) = f_3(x) = 0. \label{case2}
\end{eqnarray}
 This mapping turns Eq. (\ref{eq:dirac}) into
\begin{eqnarray}
e^{-i\,F_2(x)\sigma_y-i\,F_4(x)}i\dot{\phi} =e^{i\,F_2(x)\sigma_y-i\,F_4(x)}\nonumber\\(-i\sigma_x \phi '\,+\sigma_z\,m\phi).\label{eq:resultdirac}
\end{eqnarray}
Although we cannot eliminate the exponentials in both sides of the above equation as in previous cases,  Eq.~(\ref{eq:resultdirac}) is an approximate Dirac equation  in  regions  where $F_2(x)\simeq 0$. Notice that this does not necessarily entail $f_2(x)\simeq 0$, since the relation expressed in Eq.~(\ref{case2}) shows that we have a condition between $F_2'(x)$ and $f_2(x)$.

\subsection{Massless system} 
We show now how to include a Dirac mass term into a massless dynamics using previous techniques. For massless particles the Majorana and Dirac equations are equivalent and can be written as
\begin{equation}\label{massless}
i\dot{\psi} = -i\sigma_x\psi' + V\psi.
\end{equation}
We consider now that the potential in the above equation has the form  $V(x) = f_2(x) \sigma_z$. Through the transformation $\psi = e^{-iF_2(x)\sigma_y} \phi$ and imposing the constraint $F'_2(x) =-i f_2(x)$, Eq.~(\ref{massless}) becomes
\begin{equation}
e^{-iF_2(x) \sigma_y} i\dot{\phi}= -e^ {iF_2(x)\sigma_y}\, i\sigma_x \phi'. \label{eq:result2}
\end{equation}
In the cases in which $F_2(x)\simeq 0$, we can cancel the exponentials in both sides of Eq.~(\ref{eq:result2}),  obtaining a free dynamics for the wave function $\phi$. This condition is fulfilled by $F_2(x) = -imx$ when $x\simeq0$, such that the transformation $\psi = e^{-mx\sigma_y}\phi$ relates the dynamics between massive and massless particles. That is, the dynamics of  Eq.~(\ref{massless}) with $V=\sigma_z m$ ($f_2 = m$) is mapped onto the free dynamics in a small region near  the coordinate origin.

\subsection{Examples}
We illustrate now our method with some examples.  First, we will consider the 1D Dirac  and Majorana equations with a linear potential,
\begin{equation}
V(x)=g\,x, \label{eq:potential}
\end{equation}
and the transformation,
 \begin{equation}
\psi(x,t)=e^{-\frac{i\,g\,x^2\,\sigma_x}{2}}\phi(x,t).\label{eq:change}
\end{equation}
\begin{figure*}[t!]
  \includegraphics[width=0.87\linewidth]{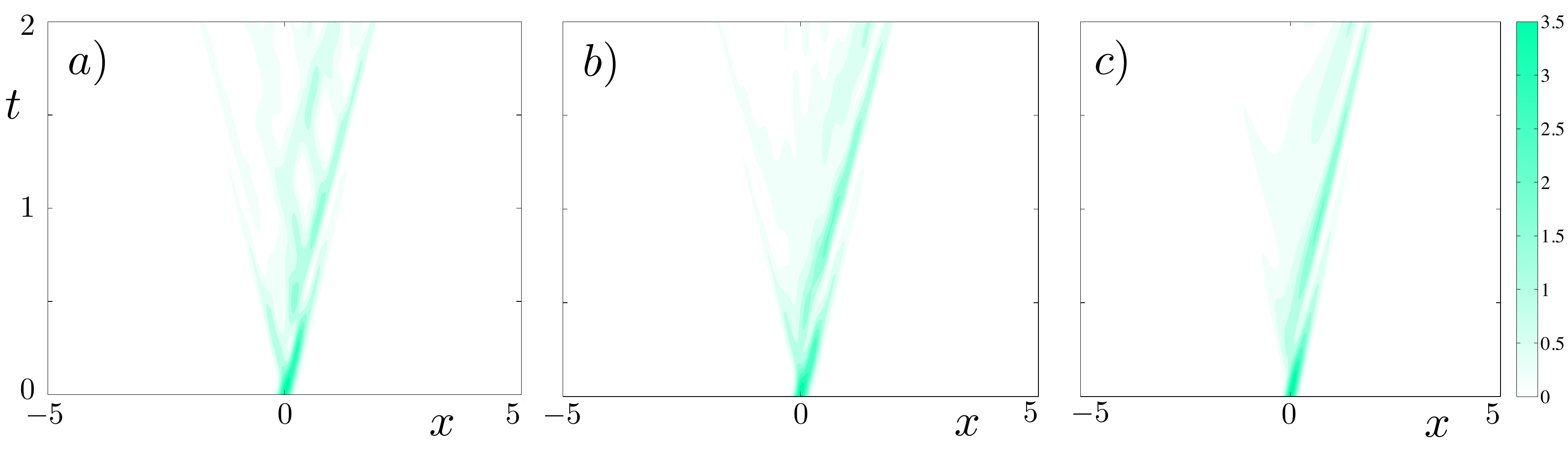}
  \caption{ (Color online)  a) Evolution of the probability density of a Dirac fermionic  wave packet under  the potential of Eq. (\ref{eq:potential2}), b) evolution of the corresponding state with the transformation in Eq. (\ref{eq:change2}) under a free Dirac equation, c) evolution for the free Dirac case without transformation. For this example, we have chosen  $m=4$, $g=2$, and $\lambda=15$.}
  \label{fig:fig37}
\end{figure*}

In this case, $\phi$ satisfies the free Majorana equation and  $|\psi|^2=|\phi|^2$, such that a solution of the Majorana equation with potential has the same probability density of a solution of the free Majorana equation. The same transformation does not work in the Dirac case, shedding light on the different behavior of Majorana and Dirac particles under such potentials \cite{MajoranaSimulation:2010}.

In Fig. \ref{fig:fig37}, we show that our method works as a good approximation in a certain region of space in the case of the Dirac equation with a potential,
\begin{equation}
V(x)=g\,\cos(\lambda x)\,\sigma_z, \label{eq:potential2}
\end{equation}
and the corresponding transformation,
 \begin{equation}
\psi(x,t)=e^{-\frac{\,g\,\sin(\lambda x)\,\sigma_y}{\lambda}}\phi(x,t).\label{eq:change2}
\end{equation}
Notice that the free dynamics (Fig. \ref{fig:fig37}b) of the transformed state is very similar to the dynamics under the potential of the untransformed state (Fig. \ref{fig:fig37}a) even in regions of space where the effect of the potential is non-trivial -as can be seen by comparing Fig. \ref{fig:fig37}a with Fig. \ref{fig:fig37}c, where the free evolution of the untransformed state is plotted.
\section{Bipartite systems.}
Our previous results can be extended to two-particle systems. For instance, it has been shown that a Lorentz-invariant two-body Dirac equation with an oscillator-like interaction can be written in the center of mass reference frame \cite{moshinski}. In 1D and our particular representation, we have
\begin{equation}
i\dot{\psi}=-\frac{i}{\sqrt2}\,(\alpha_{1}-\alpha_{2}) (\psi '+m\omega x\beta_{12}\psi)+(\beta_1+\beta_2)m\,\psi, \label{eq:dirac-osc2}
\end{equation}
with $\alpha_1=\sigma_x\otimes\mathbf{1}$, $\alpha_2=\mathbf{1}\otimes\sigma_x$, $\beta_1=\sigma_z\otimes\mathbf{1}$, $\beta_2=\mathbf{1}\otimes\sigma_z$, $\beta_{12}=\sigma_y\otimes\sigma_y$, and $x=(x_1-x_2)/\sqrt2$.
Thus, the corresponding two-body Majorana oscillator equation is,
\begin{equation}
i\dot{\psi}=-\frac{i}{\sqrt2}\,(\alpha_{1}-\alpha_{2}) (\psi '+m\omega x\beta_{12}\psi)-i(\hat{\beta}_1+\hat{\beta_2})m\,\psi^*, \label{eq:majorana-osc2}
\end{equation}
with $\hat{\beta_1}=\sigma_y\otimes\mathbf{1}$ and $\hat{\beta_2}=\mathbf{1}\otimes\sigma_y$. With the techniques explained above, we find that the mapping
 \begin{equation}
\psi(x,t)=e^{-\frac{\,m\,\omega x^2\,\beta_{12}}{2}}\phi(x,t),\label{eq:change3}
\end{equation}
transforms Eq. (\ref{eq:majorana-osc2}) into a free two-body Majorana equation near the coordinate origin,
\begin{equation}
e^{-\frac{\,m\,\omega x^2\,\beta_{12}}{2}}i\dot{\phi}=e^{\frac{\,m\,\omega x^2\,\beta_{12}}{2}}[-\frac{i}{\sqrt2}\,(\alpha_{1}-\alpha_{2}) \phi '-i(\hat{\beta}_1+\hat{\beta_2})m\,\phi^*].\label{eq:majorana-free2}
\end{equation}
Thus, in this case the situation is  similar to the one-particle example of Fig. \ref{fig:fig37}, i.e., Eq. (\ref{eq:majorana-osc2}) can be simulated with good approximation with a two-body free Majorana equation in a certain region of space. Additionally, the probability density is not conserved under the transformation. Interestingly, the same does not work for the two-body Dirac oscillator.

\section{Implementation}
In order to realize this kind of protocol, the most involved part is the creation of the initial free-evolving state by the  static transformation that encodes the potential dynamics onto free-evolving states. This can be much easier than the implementation of the corresponding potential in many different cases~\cite{LeibfriedEtAl,LawEberly}. For example, in the case of the transformation in Eq.~(\ref{eq:change}), this can be implemented in trapped ions by a straightforward Jaynes-Cummings plus anti-Jaynes-Cummings interaction in the dispersive limit, through the Hamiltonian $H=\hbar\eta\Omega(\sigma_+e^{i\delta t}+\sigma_-e^{-i\delta t})(a+a^\dag)\simeq \hbar[(\eta\Omega)^2/\delta]\sigma_z(a+a^\dag)^2\propto x^2\sigma_z$, plus a local rotation to change from the $\sigma_z$ to the $\sigma_x$ basis. Here $\eta$ is the Lamb-Dicke parameter, $\Omega$ is the laser Rabi frequency, $\sigma_{+,-,x,z}$ are the corresponding spin operators, and $a$, $a^\dag$ are the phononic annihilation and creation operators.  For general transformations, i.e., arbitrary functions of $x$ coordinate times Pauli matrices, one may generate a unitary operator $U$ that implements an arbitrary continuous-variable state by concatenated application of Jaynes-Cummings plus carrier interactions~\cite{LawEberly}. To obtain the coupling to the spin degree of freedom, standard quantum conditional logic techniques for implementing the corresponding controlled-$U$ gate may be used~\cite{NielsenChuang}. In general, with our formalism, the local transformation can be non-unitary, but always will map a pure state onto a pure state, given that it is just a matrix operator acting upon a vector state. This can be always realized by normalizing the final state and checking for the appropriate unitary transformation that connects initial and normalized final states. A normalization constant does not introduce any modification to the protocol given that all the considered equations are linear. In order to implement this kind of local transformation upon a specific implementation like trapped ions, coupling between spin and motional degrees of freedom is provided by red and blue sideband laser pulses. A wide variety of static transformations can be implemented in this way~\cite{LeibfriedEtAl}.

\section{Conclusions}
We have shown that Majorana and Dirac dynamics with a certain class of potentials can be simulated with the corresponding free dynamics by means of a static mapping between free states and states under the action of a potential. The Dirac and Majorana potentials for which this procedure is valid are of different nature, illustrating the different behavior of Majorana and Dirac dynamics \cite{MajoranaSimulation:2010}. We also obtain that the probability densities of free and non-free states coincide for arbitrary scalar potentials $V(x)=f_1(x)$ in the Majorana equation. This implies that Majorana particles behave under this kind of potential as if they were free.  In other cases, the method works as a good approximation in certain regions of space or parameter ranges. In particular, for massless Dirac systems a mass-like potential can be simulated in this way. We have extended our results to two-particle interacting systems.

The authors would like to thank I. L. Egusquiza for his useful comments. The authors acknowledge funding from Basque Government  grants BFI08.211, IT559-10, and IT472-10; Spanish MICINN FIS2008-05705, FIS2009-10061, and FIS2009-12773-C02-01; QUITEMAD; EC Marie-Curie program; UPV/EHU UFI 11/55; CCQED, SOLID and PROMISCE European projects.

\end{document}